# Identification of Crystal Symmetry from Noisy Diffraction Patterns by A Shape Analysis and Deep Learning

Leslie Ching Ow Tiong[1†], Jeongrae Kim[1†], Sang Soo Han[1*], Donghun Kim[1*]

[1]Computational Science Research Center, Korea Institute of Science and Technology (KIST), Seoul 02792, Republic of Korea

[†]These authors contributed equally.
[*]Corresponding author: sangsoo@kist.re.kr (Sang Soo Han); donghun@kist.re.kr (Donghun Kim)

# Abstract

The robust and automated determination of crystal symmetry is of utmost importance in material characterization and analysis. Recent studies have shown that deep learning (DL) methods can effectively reveal the correlations between X-ray or electron-beam diffraction patterns and crystal symmetry. Despite their promise, most of these studies have been limited to identifying relatively few classes into which a target material may be grouped. On the other hand, the DL-based identification of crystal symmetry suffers from a drastic drop in accuracy for problems involving classification into tens or hundreds of symmetry classes (e.g., up to 230 space groups), severely limiting its practical usage. Here, we demonstrate that a combined approach of shaping diffraction patterns and implementing them in a multistream DenseNet (MSDN) substantially improves the accuracy of classification. Even with an imbalanced dataset of 108,658 individual crystals sampled from 72 space groups, our model achieves 80.2% space group classification accuracy, outperforming conventional benchmark models by 17-27 percentage points (%p). The enhancement can be largely attributed to the pattern shaping strategy, through which the subtle changes in patterns between symmetrically close crystal systems (e.g., *monoclinic* vs. *orthorhombic* or *trigonal* vs. *hexagonal*) are well differentiated. We additionally find that the novel MSDN architecture is advantageous for capturing patterns in a richer but less redundant manner relative to conventional convolutional neural networks. The newly proposed protocols in regard to both input descriptor processing and DL architecture enable accurate space group classification and thus improve the practical usage of the DL approach in crystal symmetry identification.

# Introduction

High-throughput material synthesis and characterization have been popular topics of research during the last few decades and have accelerated the discovery of novel materials [1–5]. Although various characterization methods exist, identifying the crystal symmetry, i.e., the way the atoms are arranged in space, is inarguably the first and most important process in material characterization. This is because the crystallographic structure of a material plays an important role in determining the material properties (structure-property relationship) [6,7]. For a concrete example, consider the magnetism of iron: bcc Fe is ferromagnetic, while fcc Fe shows paramagnetic behaviors [8]. The most effective way to classify crystal symmetries is to find the group representing all transformations under which a system is invariant, namely, its space group. In three dimensions, there are 230 distinct types of space groups when chiral copies are considered [9–11]; these space groups are formed from the combinations of the 32 point groups with the 14 Bravais lattices [12]. Manually determining the space group to which a target material belongs is a tedious and highly inefficient task due to the brute-force nature of the search algorithms, which are based on matching diffraction patterns to those in a database, such as the Crystallography Open Database or the Inorganic Crystal Structure Database [6,13–17]. Thus, there is a strong and timely need for robust and automated assessment tools for crystal symmetry determination.

Techniques based on X-ray and electron-beam diffraction are the most related to the identification of crystal symmetries. The latest generation of tools for diffraction experiment allows the simultaneous collection of large volumes of data [18,19], the handling of which calls for big data techniques and machine-learning-based approaches. Several recent works have introduced regression models or deep learning (DL) models for material characterization. Liu *et al.* [20] refined atomic pair distribution functions in a convolutional neural network (CNN) to classify space groups. For similar purposes, Park *et al.* [21], Vecsei *et al.* [22], Wang *et al.* [23] and Oviedo *et al.* [24] used powder X-ray diffraction (XRD) 1D curves, for which information such as peak positions, intensities, and full widths at half maximum (FWHM) are mainly treated as the key input descriptors. In addition, Ziletti *et al.* [25] (in a parent work of this study), Aguiar *et* *al.* [26], Kaufmann *et al.* [27], and Ziatdinov *et al.* [28] developed DL models by extracting features from electron-beam based 2D diffraction patterns. These studies clearly show that DL methods can effectively reveal correlations between diffraction data and crystal symmetry. Despite their promise, however, most of these studies have been limited to identifying relatively few classes or crystal systems into which a material can be grouped. DL-based methods of crystal structure determination work perfectly for problems with a small number of symmetry classes (fewer than 10); however, they suffer from a drastic drop in accuracy for more difficult problems involving classification into tens or hundreds of symmetry classes (e.g. up to 230 space groups), severely limiting their practical usage. A DL model that is capable of identifying hundreds of classes with a sufficiently high accuracy will be needed to realize a robust, automated, and ultimately self-driving microscopy system or laboratory [29–31].

In this work, considering the limitations imposed by the spotty and noisy distributions of raw diffraction patterns (DPs), we propose a solution, namely, shaped DPs in a multistream DenseNet (MSDN). Our new method greatly enhances the accuracy of space group classification. Even for an imbalanced dataset of 108,658 crystals sampled from 72 space groups, the model achieves 80.2%, exceeding the performance of benchmark methods by 17-27 percentage points (%p). We find that the shaping strategy enhances the uniqueness of the raw DPs; hence, even small observable differences between raw images of symmetrically close crystal systems (e.g., *monoclinic* vs. *orthorhombic* or *trigonal* vs. *hexagonal*) become pronounced. In addition, the introduction of the MSDN allows the patterns to be captured in a richer but less redundant manner than is possible in a standard CNN. Owing to their substantial performance enhancements, our proposed methodological protocols show promise for improving the practical usage of DL approaches in crystal symmetry determination.

# Results

**Shaped diffraction patterns in a multistream DenseNet**

Raw DPs are spotty and noisy and, thus, difficult to learn from. To enhance the capabilities of DL, we propose two ideas: one is to shape the DPs, and the other is to implement them in a multistream DL network (Figure 1). The former strategy is to refine the raw DPs by selectively connecting nodes, which transforms them into shaped DPs. One can expect three possible benefits from shaped DPs: (1) the learning objective becomes more solid; (2) by controlling the shaping criteria, it is possible to maximize the uniqueness of each diffraction pattern; and (3) the added lines may amplify critical information such as lattice parameters (length, angles, etc.). We hypothesize that these benefits will result in improved deep learning of crystal symmetries.

Shaped DPs are produced as follows. First, raw DPs are collected from three orthogonal zone axes (the $x$-, $y$-, and $z$-axes) in the Condor software with an incident beam wavelength λ of $3.5 \times 10^{-12}$ m [32]. In Figure 1b, let $\mathbf{R}_* = \{N_{*,1}, N_{*,2}, \cdots, N_{*,n}\}$ be the raw DPs, where the $N_*$ represent each node composed of multiple pixels, $n$ is the number of nodes, and * denotes each axis. The distances between node pairs are then calculated, i.e., $dist_{N_*} = d(N_{*,i}, N_{*,j})$, where $N_{*,i}$ and $N_{*,j}$ are two arbitrary nodes and $d(\cdot)$ is the Euclidean distance function. We draw interpolated lines only for node pairs with a distance smaller than a certain threshold, i.e., $1.7 \times \min(dist_{N_*})$. The prefactor 1.7 was determined after extensive tests: the shapes become too complex with a larger threshold value, whereas the shapes are not clearly formed with a smaller threshold value. The colors R, G, and B are used for lines in images of the $x$-, $y$-, and $z$-axes. Thus, the shaped DP, or $\mathbf{S}_*$, is calculated as $\mathbf{R}_* + \sum lineplot(N_{*,i}, N_{*,j})$, where the sum $\sum$ is taken over the selected node pairs and $lineplot(\cdot)$ is the interpolation function. As shown in the scheme of the DP shaping process (Figure 1b), the $lineplot(\cdot)$ function is dependent on the node sizes; as a result, the line thickness will differ for different node pairs. Additional information related to the DP shaping protocols is provided in the Methods and in Supplementary Figure 1. As seen in the examples from several space groups presented in Figure 1b and Supplementary Figure 2, the shaped DPs are more solid and much less noisy than the raw versions. The resulting shapes comprise composition information that describes the particular regions of interest that are useful for representing DPs in more unique manners.

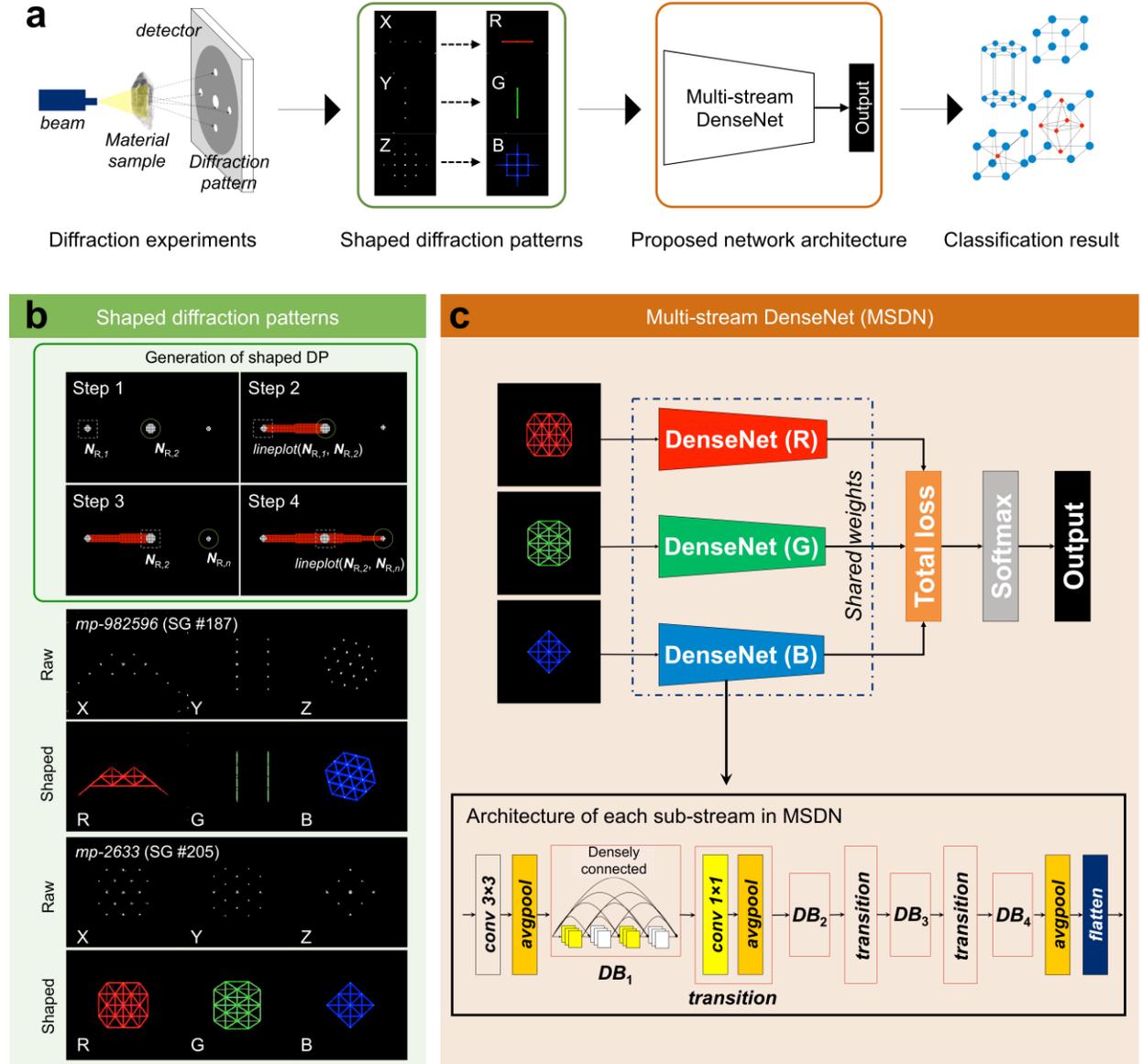

**Figure 1. Shaped diffraction patterns in an MSDN. a,** A scheme that describes the automated determination of crystal symmetry based on diffraction experiments. **b,** A scheme describing the generation process for shaped DPs as well as two exemplary results from space groups #187 and #205. Note that in the generation scheme, the line thickness depends on the node size, which makes the shapes more unique. **c,** The network architecture of the MSDN.

For the further processing of multiple inputs (DPs collected from the three zone axes), we propose a novel multistream network, namely, an MSDN, as shown in Figure 1c. In the MSDN, three substream

DenseNets are applied in parallel to each shaped DP; these DenseNets share all of their parameters (weights $W$ and biases $b$). The idea of sharing parameters is warranted by the consistent learning process for all three shaped DPs ($S_R$, $S_G$, and $S_B$). This imposes prior knowledge that the inputs to each substream are processed concurrently by the network, which substantially reduces the number of parameters in the MSDN. In addition, the MSDN utilizes the design concept of DenseNet [33], in which all layers are densely connected (Figure 1c); in contrast, in a standard CNN, the features in each *conv* layer are used as input to the next layer without communication. The superior performance of DenseNets over standard CNNs has been previously reported in the field of image learning and classification [33–35]. Likewise, in the present study on the processing of DP images, the proposed MSDN is expected to create rich patterns while maintaining a low complexity of information, thus enabling better classification performance.

The MSDN concurrently accepts and processes shaped DPs, i.e., $S_R$, $S_G$, and $S_B$, to extract a better feature representation from each substream for space group classification. Specifically, each layer in each DenseNet receives the inputs from all preceding layers and passes its features to all subsequent layers, meaning that the final output layer has direct supervision over every single layer. As a result, the network offers stronger feature propagation for the extraction of collective knowledge in the inference process. Regarding the network configuration, the MSDN used in this study consists of four dense-block (*DB*) layers and three transition layers in each substream network, as shown in Figure 1c and Supplementary Table 1.

**Dataset**

A large-scale collection of diffraction patterns for 108,658 materials sampled from 72 space groups was acquired. These 72 space groups (out of a total of 230) were selected based on the criterion that each group should be represented by at least 295 materials in the Materials Project (MP) library [36], as shown in Figure 2a. There are too few materials (mostly <100) available for the remaining space groups in the MP library, which were therefore excluded for DL training and testing. The selected space groups include 2 *triclinic*, 12 *monoclinic*, 22 *orthorhombic*, 13 *tetragonal*, 6 *trigonal*, 8 *hexagonal*, and 9 *cubic* crystal systems. Because we downloaded the full list of materials for each space group, the dataset is highly imbalanced, ranging from 295 materials for space group #223 to 8,700 materials for space group #14. For the following DL experiments on space group classification, we constructed datasets consisting of 8, 20, 49, and 72 space groups (SGs), as shown in Figure 2b. The number of materials in each space group is tabulated in Supplementary Table 2.

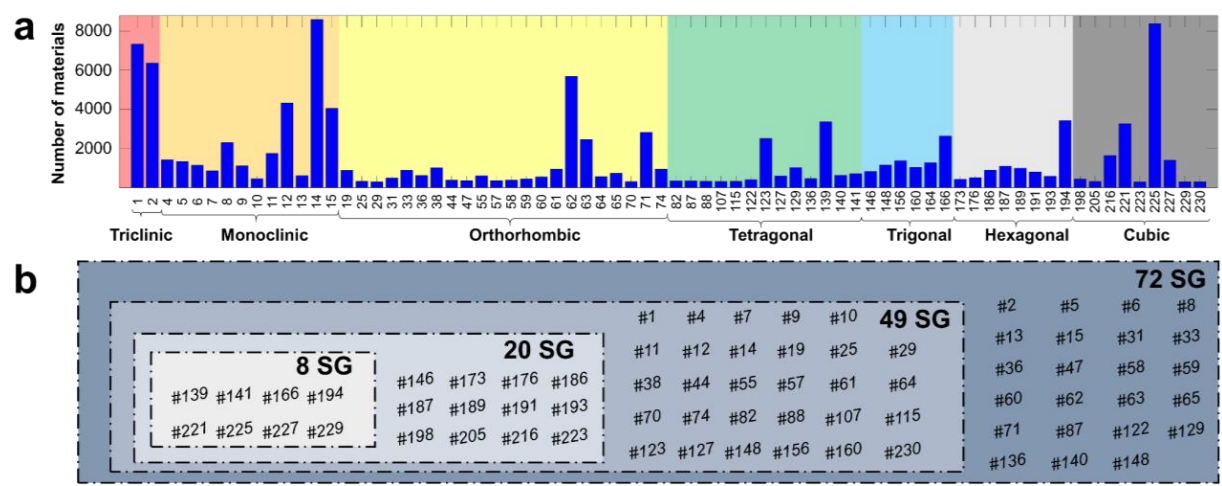

**Figure 2. Population distribution of the diffraction pattern dataset. a,** The number of materials in each space group, along with the crystal system information. The background colors represent seven types of crystal systems: *triclinic* in red, *monoclinic* in orange, *orthorhombic* in yellow, *tetragonal* in green, *trigonal* in blue, *hexagonal* in light gray, and *cubic* in dark gray. **b,** The usage of our dataset for the experiments.

**Classification experiments with varying numbers of space groups**

We conducted DL experiments to study the classification of space groups (Figure 3). To evaluate the impact of our strategy (shaped DPs in an MSDN), we performed comparisons with other benchmark models, i.e., spot DPs in AlexNet [37], DenseNet [33], ResNet [38], and VGGNet [39]. Spot DPs, which were originally proposed in the work of Ziletti *et al.* [25], are the superimposed version of the raw DPs from R/G/B color channels. See the scheme in Supplementary Figure 3 for an exemplary illustration of spot

DPs. The key parameter in our experiments was the number of space groups into which materials could be classified; we considered 8, 20, 49, and 72 (Figure 2b). In each case, the dataset was divided into 80% of the data for learning (training and validation) and 20% of the data for testing, with no overlap. In Figure 3a, to begin with the smallest-scale dataset (with 8 SGs), both our approach and the other benchmark models work excellently: ours shows 99.5% accuracy, while the others also achieve accuracies of above 94.5%. Notably, we have well reproduced the results of the state-of-the-art work of Ziletti *et al.* (over 99% for 8 SGs) [25], which indicates that our experiments are reliable.

Proceeding to more difficult problems, i.e., larger-scale datasets (20, 49, and 72 SGs), we observe that our strategy of shaped DPs in an MSDN performs substantially better than the benchmark models. In Figure 3a, our method achieves excellent top-1 classification accuracies of 99.5%, 93.0%, 84.4% and 80.2% for the 8 SG, 20 SG, 49 SG and 72 SG datasets, respectively. On the other hand, the other models based on spot DPs considerably underperform: even the leading model among the benchmarks (spot DPs in Ziletti *et al.*'s network) exhibits an accuracy of below 63% for the 72 SG dataset. This result proves the relatively high tolerance of our model to an increasing number of space groups for classification, which is a critical requirement for its practical usage. We additionally measured the performance achieved with shaped DPs in a multistream VGGNet (MSVGG) in order to distinguish the contributions from the "shaped DP" and "MSDN" aspects of the proposed strategy. For the case of the 72 SG dataset, the total enhancement of 17 %p can be divided into a 10 %p contribution from the shaped DPs and the remaining 7 %p of the contribution from the MSDN, confirming that both strategies play critical roles.

Unlike in Figure 3a, in which only the top-1 classification performance is considered, the top-$k$ ($k$=1−5) ranking accuracy is presented in Figure 3b-3e (for the 8, 20, 49, and 72 SG datasets, respectively). We observe that for all cases, our strategy of shaped DPs in an MSDN performs the best regardless of the $k$ value, followed by shaped DPs in an MSVGG. This once again confirms the superiority of shaped DPs over the conventional spot DPs as the descriptors used for crystal symmetry determination. For the smaller datasets (8 and 20 SGs), the classification is almost perfect (accuracy>99%) even at the top-2 ranking. For the larger datasets (49 and 72 SGs), the accuracy remains above 95% at the top-4 ranking (49 SG dataset) or the top-5 ranking (72 SG dataset).

The more challenging task of classification on an untrained space was also addressed in testing (Figure 3f). This task arises when the sample being tested does not fall into any of the space groups on which the classifier was previously trained. In this experiment, for testing purposes, we randomly sampled 3,052 materials from 30 additional space groups, which had no overlap with the aforementioned 72 SGs. A list of these 30 SGs is provided in Supplementary Table 3. These 3,052 material samples were divided into a reference set (50%) and a test set (50%). Next, classification was performed by measuring the cosine similarity distance between the reference set and each tested material. Details of the similarity distance calculation can be found in the Methods. Surprisingly, our network achieved a top-1 classification accuracy of 70.2% and reached 87.5% at the top-5 ranking. These accuracy values are impressively high, given that the tested materials belonged to SGs that were never considered in training. The observed generalizability of our model is likely to be beneficial in real situations in which the tested materials are not part of the training space.

**Classification results for individual space groups**

We investigated the classification results for individual space groups. Only the 49 SG and 72 SG cases were analyzed (Figure 4a and Figure 4b). An interesting observation for both benchmarks and our model is that the accuracy is generally higher for SGs in high-symmetry crystal systems. The classification process tends to work much better for *cubic*/*hexagonal*/*trigonal* systems than for *monoclinic*/*orthorhombic* ones. *Triclinic* systems are an exception, largely due to the insufficient number of materials belonging to these systems. In Figure 4c and Figure 4d, while the benchmarks show the highest accuracy for *cubic* systems, the accuracy of our model is the highest for *trigonal* and *hexagonal* systems rather than cubic systems. In particular, for the 49 SG dataset, it is observed that for all space groups corresponding to *trigonal* and *hexagonal* systems (#146−#194), the classification accuracy is excellent, being over 90%.

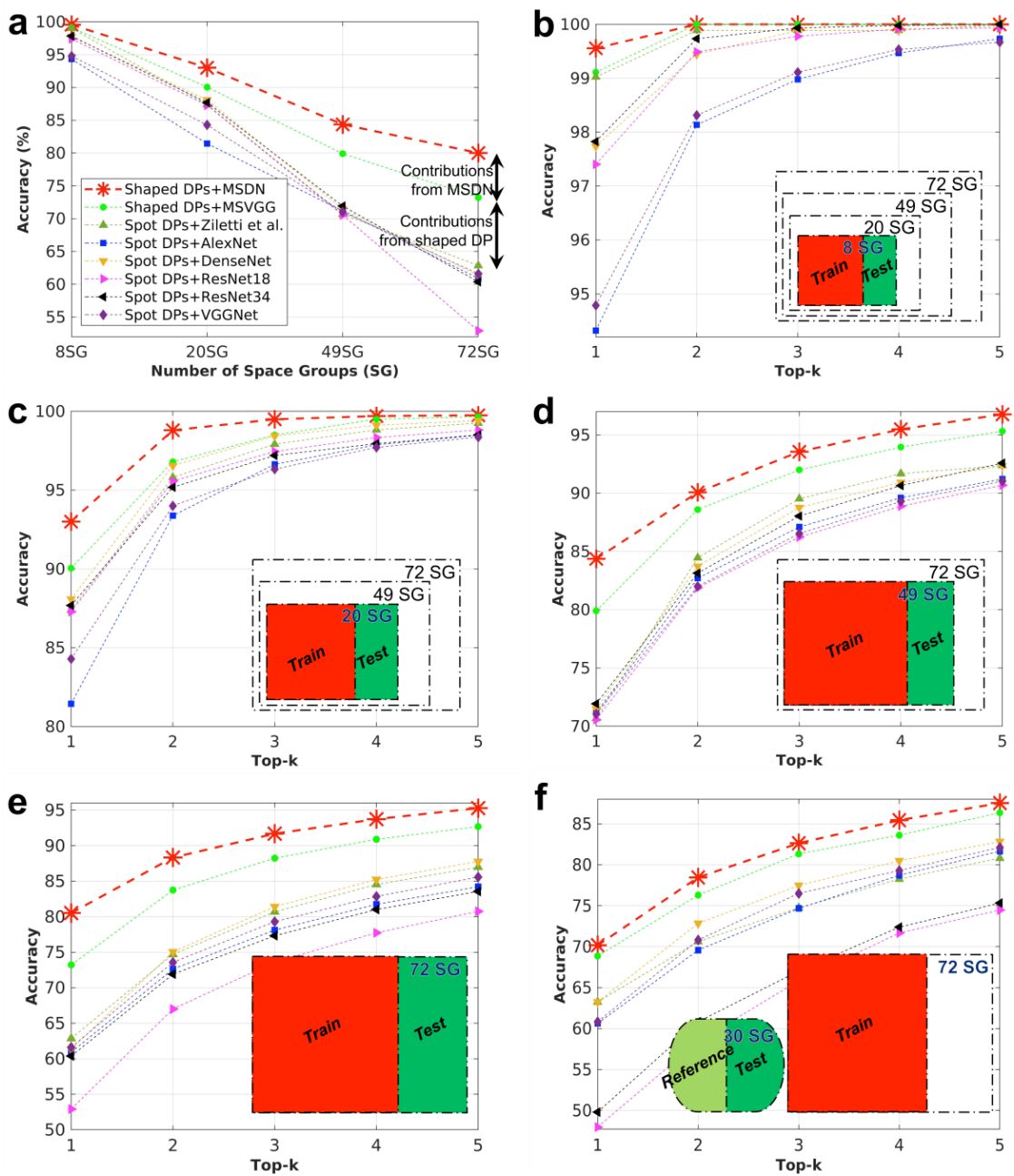

**Figure 3. Space group classification performance**. **a,** Top-1 accuracy as a function of the number of space groups for classification. **b-e,** Top-*k* accuracies for the datasets consisting of 8 SGs (**b**), 20 SGs (**c**), 49 SGs (**d**), and 72 SGs (**e**). **f,** Top-*k* accuracies for testing an untrained space with an additional 30 SGs. The top-*k* accuracy refers to the percentage of cases in which the correct class label appears among the top-*k* probabilities.

The accuracy improvements in our model over the benchmarks appear to be universal for most SGs. To identify the source of these improvements, we now decompose the contributions for each crystal system (Figure 4c and Figure 4d). The model named spot DPs+Ziletti *et al.* is selected as the representative benchmark here due to its relatively high performance. Triclinic systems are excluded from the analysis due to the statistically insufficient number of materials. The enhancements in accuracy are ranked as follows: *trigonal* (24.1 %p) > *monoclinic* (19.7 %p) > *hexagonal* (18.1 %p) ≈ *tetragonal* (18.11 %p) > *orthorhombic* (13.7 %p) > *cubic* (4.8 %p), where the values in parentheses are the average values for the 49 and 72 SG datasets. The contribution for *cubic* systems is much smaller than those for the other crystal systems.

Next, we focus on further characterizing the incorrect classifications obtained from the benchmark (spot DPs+Ziletti *et al.*) and our model (shaped DPs+MSDN). In Figure 4e and 4f, for instance, the [*monoclinic*, *orthorhombic*] coordinate in the matrices represents the materials belonging to an SG corresponding to a *monoclinic* system that were incorrectly classified as belonging to an *orthorhombic* system. In the comparisons between the benchmark and our model, the most prominent changes are observed in two areas, i.e., the *monoclinic*/*orthorhombic* and *trigonal*/*hexagonal* pairs. This indicates that the benchmark model often finds it difficult to correctly classify SGs corresponding to *monoclinic* vs. *orthorhombic* systems or to *trigonal* vs. *hexagonal* systems, whereas our model performs much better in resolving this confusion. We speculate that such confusion may occur mainly between symmetrically close crystal systems. For instance, *monoclinic* and *orthorhombic* systems are very close in terms of lattice symmetry, differing only in the lattice angle requirements (90° angle requirements). Therefore, similar spot distributions in spot DPs can possibly arise even from materials from different crystal systems, which may undermine the performance of spot-DP-based benchmark models.

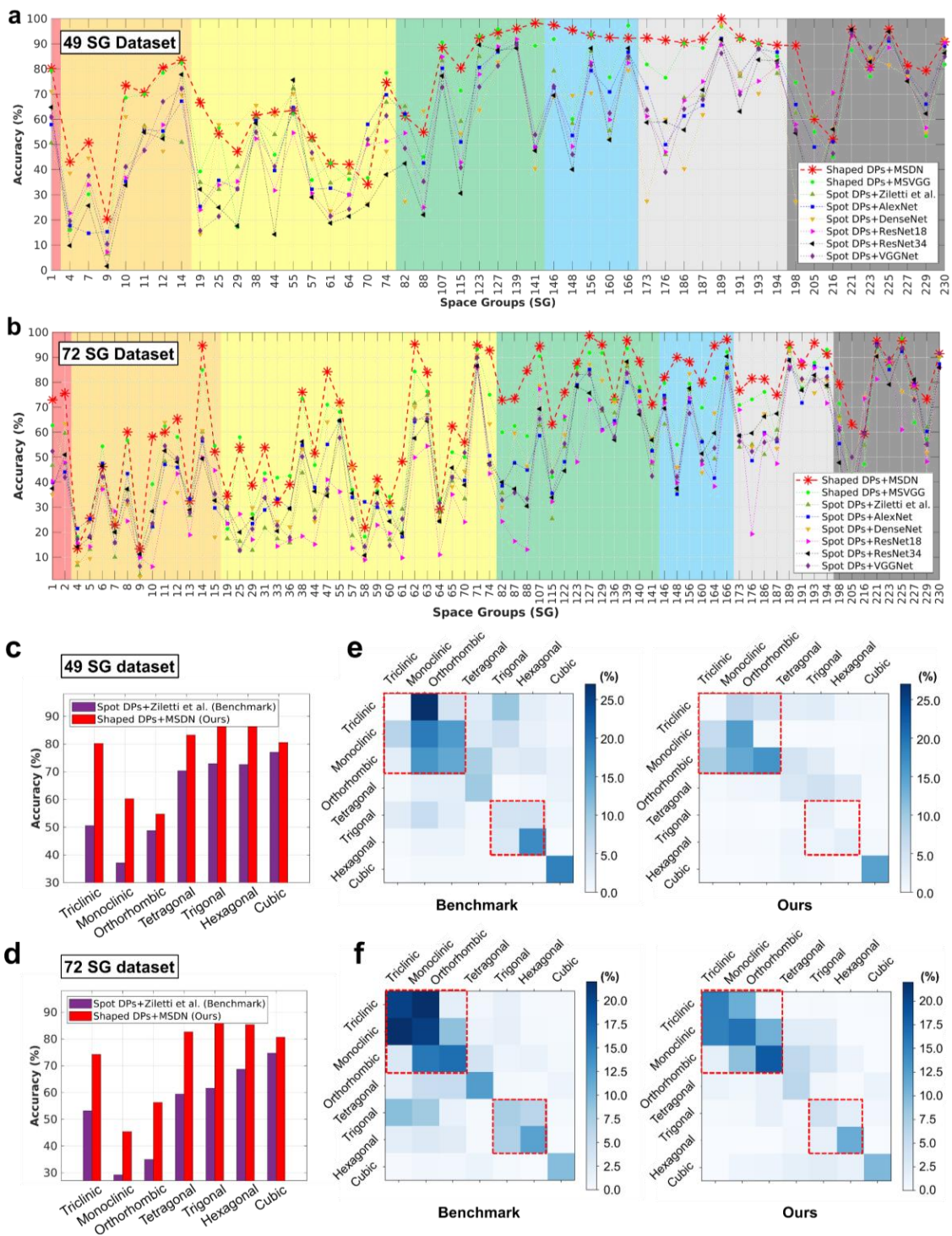

**Figure 4. Decomposition analysis to identify the origins of the performance improvement. a-b,** Classification results for individual space groups from the 49 SG (**a**) and 72 SG (**b**) datasets. The background colors represent the seven types of crystal systems, as in Figure 2a. **c-d,** Average classification accuracy by crystal system type for the 49 SG (**c**) and 72 SG (**d**) datasets. **e-f,** Matrices showing the distribution rates (%) of incorrect predictions for the 49 SG (**e**) and 72 SG (**f**) datasets. If the rate is, for example, 20% for the [*monoclinic*, *orthorhombic*] coordinate in a matrix, this means that 20% of the materials belonging to *monoclinic* systems in our dataset are incorrectly classified as belonging to SGs corresponding to *orthorhombic* systems. Red dotted boxes highlight the regions that are considerably different between the benchmark and our model.

To further justify our observation that our model (shaped DPs+MSDN) can largely resolve the confusion between symmetrically close systems, we scrutinize the DPs of several test samples. Figure 5 shows exemplary cases in which spot DPs fail and shaped DPs succeed in yielding correct SG classifications. For the first two example pairs of *mp-1076884* (SG #1, *triclinic*) vs. *mp-6406* (SG #7, *monoclinic*) and *mp-6019* (SG #14, *monoclinic*) vs. *mp-556003* (SG #74, *orthorhombic*), the raw and spot DPs are both too similar (almost identical) to be easily differentiated. This is consistent with the powder X-ray diffraction data available in the MP library in which the peak locations and intensities are alike. However, the shaped DPs look substantially different, enabling the correct SG classification of these samples. In appearance comparisons of the shaped DPs, we find that the shaped DPs appear more symmetric for the higher-symmetry crystal system, as seen in the R-channel image for the first example pair (*triclinic* vs. *monoclinic*) and the G- and B-channel images for the second example pair (*monoclinic* vs. *orthorhombic*). The result indicates that the shape analysis can distinguish even small differences (barely observable by human eyes) in node position, size, and brightness, which are likely to be induced by the different level of lattice symmetries of crystal systems.

For the latter two example pairs of *mp-757070* (SG #166, *trigonal*) vs. *mp-1195186* (SG #176, *hexagonal*) and mp-5055 (SG #186, *hexagonal*) vs. *mp-29211* (SG #160, *trigonal*), although the raw and spot DPs do look slightly different, the benchmark models unfortunately do not predict the correct SGs for these samples. In the shaped DPs, however, these subtle differences are maximized. Notably, the distance information of adjacent node pairs, which is often related to the lattice parameters, is greatly amplified in the shaped DPs, as observed in the R and B channels of the 4$^{th}$ example pair. From these case studies, we find that the shaping strategy enhances the uniqueness of the raw DPs more than the superimposition strategy used to produce the spot DPs does; hence, even small observable differences in pattern between symmetrically close crystal systems (e.g., *monoclinic* vs. *orthorhombic* or *trigonal* vs. *hexagonal*) become pronounced.

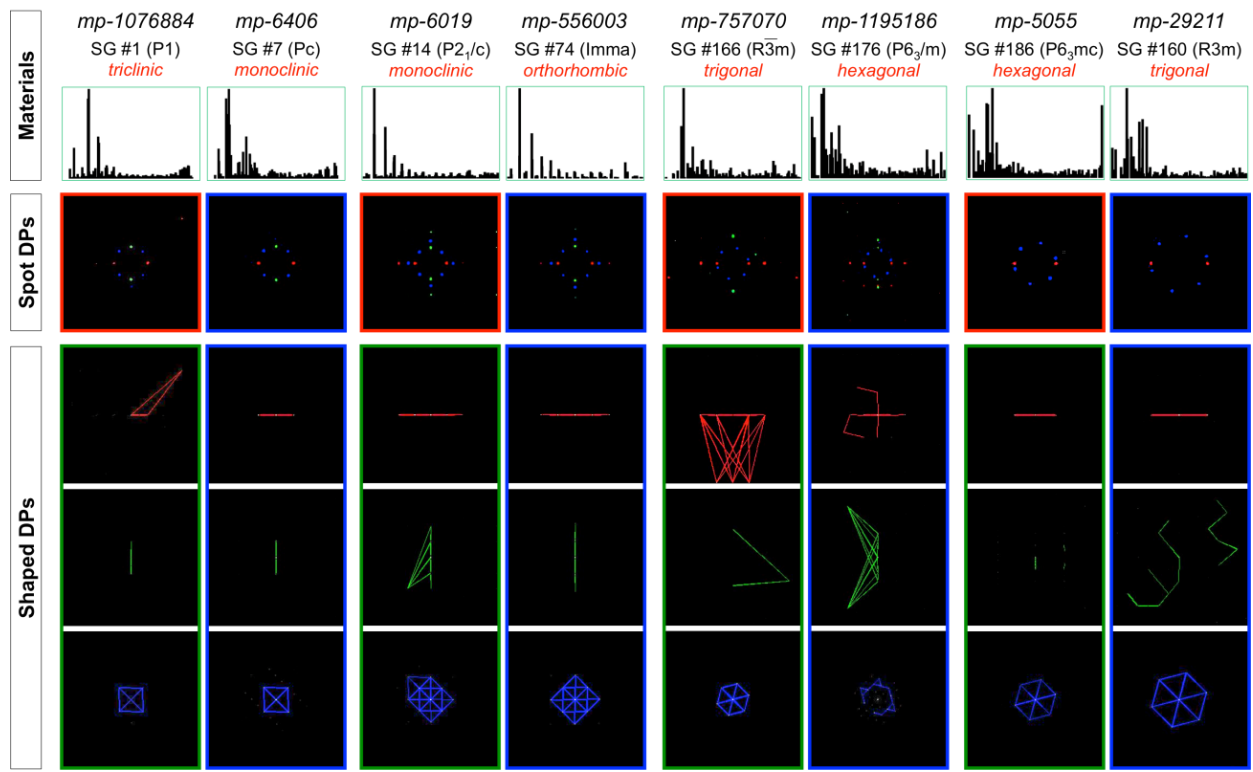

**Figure 5. Case studies in which spot DPs fail and shaped DPs succeed in yielding correct SG classifications.** The top row provides the material information of the test samples, which are available in the MP library, including the MP id, SG #, and powder X-ray diffraction data. The chemical formula of each material is as follows: *mp-1076884* ($Sr_6Ca_2Fe_7CoO_{20}$), *mp-6406* ($Na_2MgSiO_4$), *mp-6019* ($Sr_2YNbO_6$), *mp-556003* ($CaTiO_3$), *mp-757070* ($BaCaI_4$), *mp-1195186* ($RbLa_2C_6N_6ClO_6$), *mp-5055* ($Na_6MnS_4$), *mp-29211* ($V_4Cu_3S_8$). The next four rows show the spot DPs and shaped DPs of each material. The green and red boxes indicate success and failure cases, respectively, for SG classification, and the blue boxes refer to the reference data in the training set. Best viewed in an electronic version.

In addition to the shaping strategy, the MSDN architecture also contributes to the performance improvements; here, we would like to discuss the benefits of this network. Figure 6 visualizes both the *conv* layers from the MSVGG and the *DB* layers from the MSDN for selected diffraction images. Several additional examples are presented in Supplementary Figures 4 and 5. The visualization results show that the patterns captured in the MSDN are clearer, richer, and less redundant than those in the MSVGG. Indeed, several feature patterns in the MSVGG are redundant, such as those for samples A, C, and D (highlighted in the red dotted boxes), while such redundant feature patterns are not found in the MSDN.

This is likely because the MSDN reuses the features from previous layers to prevent redundancy within the network (Supplementary Figure 6).

We also compared the computational and memory efficiency of the MSVGG and MSDN. The MSDN is superior to the MSVGG in terms of both space complexity (total number of parameters) and time complexity (FLOPS: floating-point operations per second). The numbers of parameters and FLOPS are 128.85M and 515.37M, respectively, for the MSVGG, while they are much smaller at 1.54M (84 times smaller) and 5.75M (90 times smaller), respectively, for the MSDN. In fact, the number of parameters of the MSVGG is enormous because every single layer has its own weights and biases (***W*** and ***b***) to be learned. In the MSDN, this complexity is avoided by optimizing the parameters and simplifying the connectivity between layers because it is unnecessary to learn redundant feature maps. Such a large difference is possible because the MSDN can receive direct supervision for the propagation of the error signal from the preceding layers to the final layer. These comparisons indicate that DP image processing is extremely fast and efficient in our MSDN model.

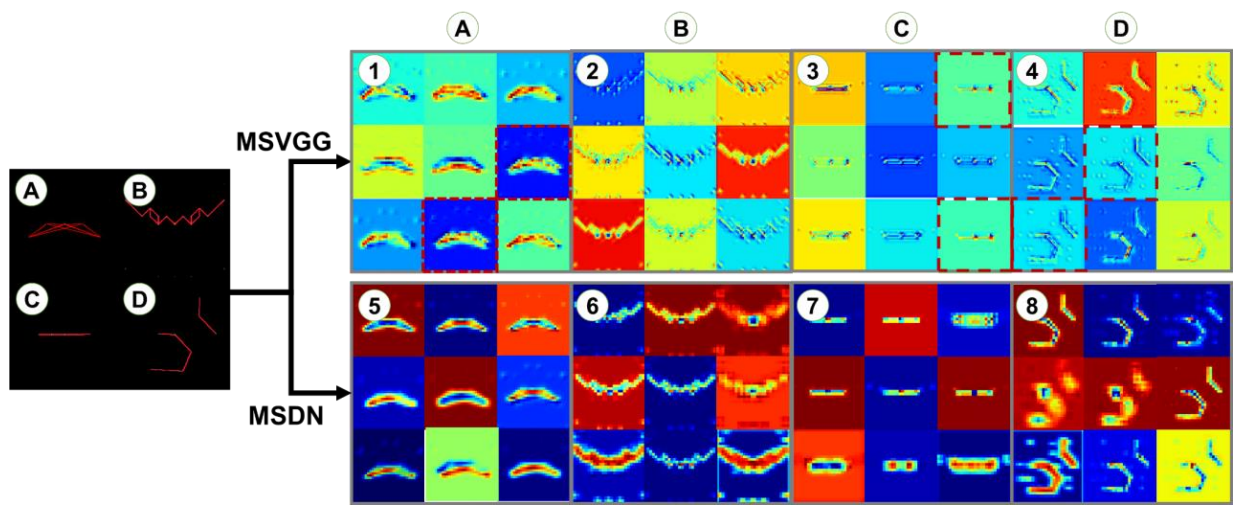

**Figure 6. Benefits of the MSDN over the MSVGG in processing DP images**. For selected exemplary diffraction images A, B, C, and D, the block layers of the MSVGG (1, 2, 3, and 4) and MSDN (5, 6, 7, and 8) are visualized. The 3$^{rd}$ *conv* block of the MSVGG and the *DB$_2$* layer of the MSDN are shown for comparison. The red dotted box indicates redundant (almost identical) feature maps. Best viewed in an electronic version.

## Discussion

It is worth discussing the limitation of the present study and next challenges. Although the combined approach of the shape analysis and MSDN architecture improves the prediction accuracy, there is still a large room for the next success. One limitation of this study is that the DL-based test was performed on 72 SGs, rather than the whole 230 SGs. This is because statistically insufficient number of materials (<200) were available for many space groups in today's MP database. Since the MP database is gradually increasing in the number of materials, the test will follow in future, covering more SGs (ideally all SGs). Another remained challenge is to explore defective structures. Defects exist everywhere in the form of grain boundaries, dislocations, voids, and local inclusions etc, and may have a large impact on macroscopic material's properties. Identifying crystal symmetry of defected materials will be the next challenge. Lastly, our study as well as most previous attemps uses the dataset of simulated diffraction patterns, rather than experimental ones, mainly due to the limited experimental data size. However, all strategies that are proven to be effective for the simulated databse should ultimately be explored for experimental database, as soon as the sufficient number of data is prepared in a consistent manner.

In summary, we propose new methodological protocols for enhanced DL-based determination of crystal symmetry, namely, shaped DPs in an MSDN. Our new methods greatly improve the SG classification accuracy. Even for an imbalanced dataset of 108,658 crystals sampled from 72 SGs, our approach achieves an accuracy of 80.2%, outperforming benchmark models based on conventional spot DPs by 17-27 %p. Both the shaped DP strategy (~10 %p) and the MSDN architecture (~7 %p) make considerable contributions to the performance improvement. The shaping strategy enhance the uniqueness of the raw DPs; hence, even small observable differences between the raw images of symmetrically close crystal systems (e.g., *monoclinic* vs. *orthorhombic* or *trigonal* vs. *hexagonal*) become pronounced in the shaped versions. We additionally find that the MSDN architecture captures the patterns in a richer but less redundant manner than is possible in a standard CNN. This work provides new protocols in regard to both input descriptor processing and the DL architecture and, as a result, enables the robust and automated classification of space groups, which we hope will facilitate the practical usage of the DL approach in crystal symmetry determination.

## Methods

**Generating and shaping diffraction patterns**

First, using the MP library [36], the coordinates of a standard conventional cell are prepared [40]. Next, these are converted into the Protein Data Bank (PDB) format to satisfy the input-feeding requirement of Condor settings. In the Condor software, a wavelength of $\lambda=3.5\times10^{-12}$ m is used for the incident beam. Three different zone axes ($x$-, $y$-, and $z$-axis) are considered. To produce the shaped DPs, we initialize the first node in $\mathbf{R}_*$, which is assigned to $N_{R,i}$, $N_{G,i}$, and $N_{B,i}$ of the dotted $i^{th}$ circle (Supplementary Figure 1). Then, we detect the neighboring $j^{th}$ node ($N_{R,j}$, $N_{G,j}$, and $N_{B,j}$) and calculate the distance between the $i^{th}$ and $j^{th}$ nodes. For each node pair with a distance smaller than a specified threshold ($1.7\times\min(dist_{N^*})$), the algorithm will plot a line between the nodes; otherwise, the algorithm will skip this step. For the line colors, red (R), green (G), and blue (B) are used for each $x$-, $y$-, and $z$-axis DP, respectively. After plotting is performed, the shaped DP outcome is created as shown in *Step K*, Supplementary Figure 1.

**Deep learning experiment**

For the DL experiments related to Figure 3a-3e, the dataset was divided into 80% of the data for learning (training and validation) and 20% for testing, with no overlap. We then divided the images in the learning set by space group for cross-validation purposes. The cross-validation procedure was designed as follows: (1) randomly shuffle the learning set; (2) split it into 10 groups; (3) take one group as the validation set and the remaining groups as the training set; (4) repeat step 3 every 100 epochs and summarize the model evaluation scores. For the testing scheme, the test set images were used to evaluate the performance of our network.

For the proposed model (shaped DPs+MSDN), we used the Adam optimizer [41] with a learning rate of $1.0\times10^{-5}$ and a weight decay and momentum of $1.0\times10^{-7}$ and 0.9, respectively. The MSDN consists of four dense-block layers and three transition layers in each substream (Figure 1c). The structure of a dense block is illustrated in Supplementary Figure 6b. Let $DB$ be a dense block with $l$ layers $H_l$, composed of *conv*, rectified linear unit (*ReLU*) and dropout [42] layers:

$$DB = H_l([x_0, x_1, \cdots, x_{l-1}]), \tag{1}$$

where $x_0 \sim x_{l-1}$ represent feature outputs and $[\cdots]$ is defined as a concatenation operator. Then, a transition layer is implemented in every block that performs $1\times1$ *conv* and *avgpool* operations. Supplementary Table 1 shows the configuration of the proposed network in detail. During training, we defined a total loss ($\ell_{\text{total}}$) function consisting of a sum of the softmax cross-entropies $\ell$ of logit vectors and their respective encoded labels, as follows:

$$\ell_{\text{total}} = \ell(F_R) + \ell(F_G) + \ell(F_B), \tag{2}$$

$$\ell(F_*) = -\sum_t^T \sum_c^C L_{tc} \log[\boldsymbol{\delta}_{\text{SG}}(F_*)_{tc}], \tag{3}$$

$$\boldsymbol{\delta}_{\text{SG}}(F_*)_{tc} = \frac{\exp^{(F_*)_{tc}}}{\sum_c^C \exp^{(F_*)_{tc}}}, \tag{4}$$

where $*$ denotes the zone axis information (one of the color R, G and B), $F$ is a flatten layer, $L$ denotes the class labels, $T$ is the number of training samples, $C$ is the number of classes, and $\boldsymbol{\delta}_{\text{SG}}(\cdot)$ is the output layer, implemented with the softmax function. The $\ell_{\text{total}}$ function provides joint supervision for the training process of the MSDN; it can robustly aggregate the descriptors from the different substreams.

For the alternative model (shaped DPs+MSVGG), we used the Adam optimizer with a learning rate of $1.0\times10^{-5}$ and a weight decay and momentum of $1.0\times10^{-7}$ and 0.9, respectively. This network consists of 24 shared *conv* layers, 15 *maxpool* layers, and 3 *fc* layers; more details of the layer configuration are provided in Supplementary Table 4. We again implemented the $\ell_{\text{total}}$ function in equation (2) to robustly aggregate the descriptors from the different substreams. For all other benchmark networks, we also used the Adam optimizer with a learning rate of $1.0 \times 10^{-4}$ and a weight decay and momentum of $1.0 \times 10^{-6}$ and 0.9, respectively.

**Score function for testing an untrained space**

For the experiments related to Figure 3f, the reference set is represented by feature vectors $V_i = \{V_{i,\text{R}}, V_{i,\text{G}}, V_{i,\text{B}}\}$ from the last embedded layer (before the output layer) in the network, where $i = 1, 2, 3, \cdots, C$ (the number of classes). A material to be tested is represented by $P = \{P_\text{R}, P_\text{G}, P_\text{B}\}$. The total score function for this test material was computed using the following sum rule:

$$\sigma_{\text{total}}(P, V_i) = \sigma(P_\text{R}, V_{i,\text{R}}) + \sigma(P_\text{G}, V_{i,\text{G}}) + \sigma(P_\text{B}, V_{i,\text{B}}), \tag{5}$$

where $\sigma(P_*, V_{i,*}) = 1 - \cos(P_*, V_{i,*})$ is defined as the cosine similarity distance and $*$ denotes the axis information (one of the colors R, G and B). To classify the test set, a new softmax function $\gamma$ was defined as follows:

$$\gamma = \max_i [\sigma_{\text{total}}(P, V_i)]. \tag{6}$$

## Data Availability

The data samples of shaped DP descriptors are shared on the following GoogleDrive link: https://drive.google.com/open?id=1l7n6khjbbUB6cRS-xwlYw5xbyOphXBn6

## Code Availability

The codes for generating shaped DPs and the pre-trained model of MSDN are available in the GitHub repository (https://github.com/tiongleslie/crystal-structure-classification). All codes are written in Python 3.7 and the architecture of MSDN is implemented using TensorFlow r1.13.

## Acknowledgments

This work was supported by the Samsung Research Funding & Incubation Center of Samsung Electronics under Project Number SRFC-MA1801-03.

## Author Contributions

D.K. and S.S.H. conceived and designed the research. L.C.O.T. and J.K. performed the research, including data collection and deep learning tasks. All authors contributed to manuscript writing.

24## Competing Interests

The authors declare no competing interests.

## Additional Information

Supplementary Information is available for this paper at https://doi.org/XX.XXXX/.

# Supplementary Information for

# Shaped Diffraction Patterns for Enhanced Deep Learning of Crystal Symmetry

Leslie Ching Ow Tiong[1†], Jeongrae Kim[1†], Sang Soo Han[1*], Donghun Kim[1*]

[1]Computational Science Research Center, Korea Institute of Science and Technology (KIST), Seoul 02792, Republic of Korea

[†]These authors contributed equally.

[*] Corresponding author: sangsoo@kist.re.kr (Sang Soo Han); donghun@kist.re.kr (Donghun Kim)

## Step 1
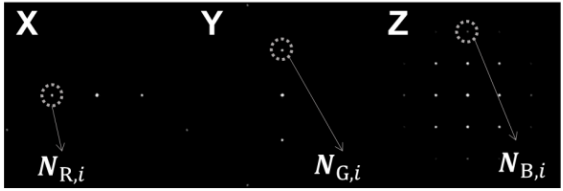

## Step 2
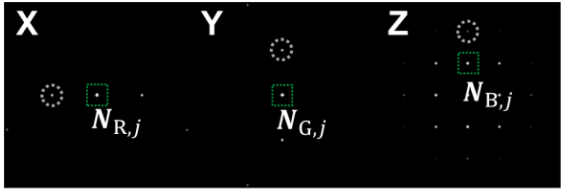

## Step 3
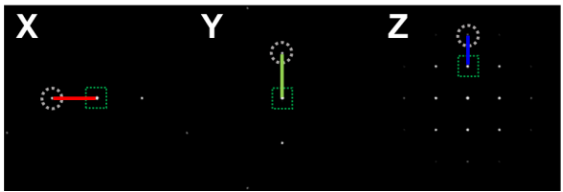

## Step 4
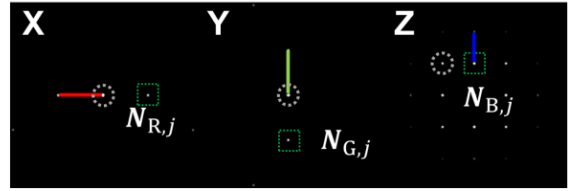

## Step 5
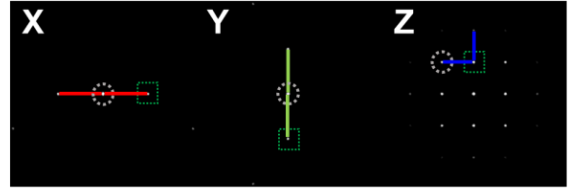

## Step K
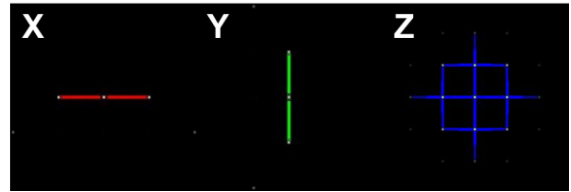

**Supplementary Figure 1**. **Step-by-step generation of shaped DP.** Each R/G/B color is used for lines in images of X/Y/Z axis.

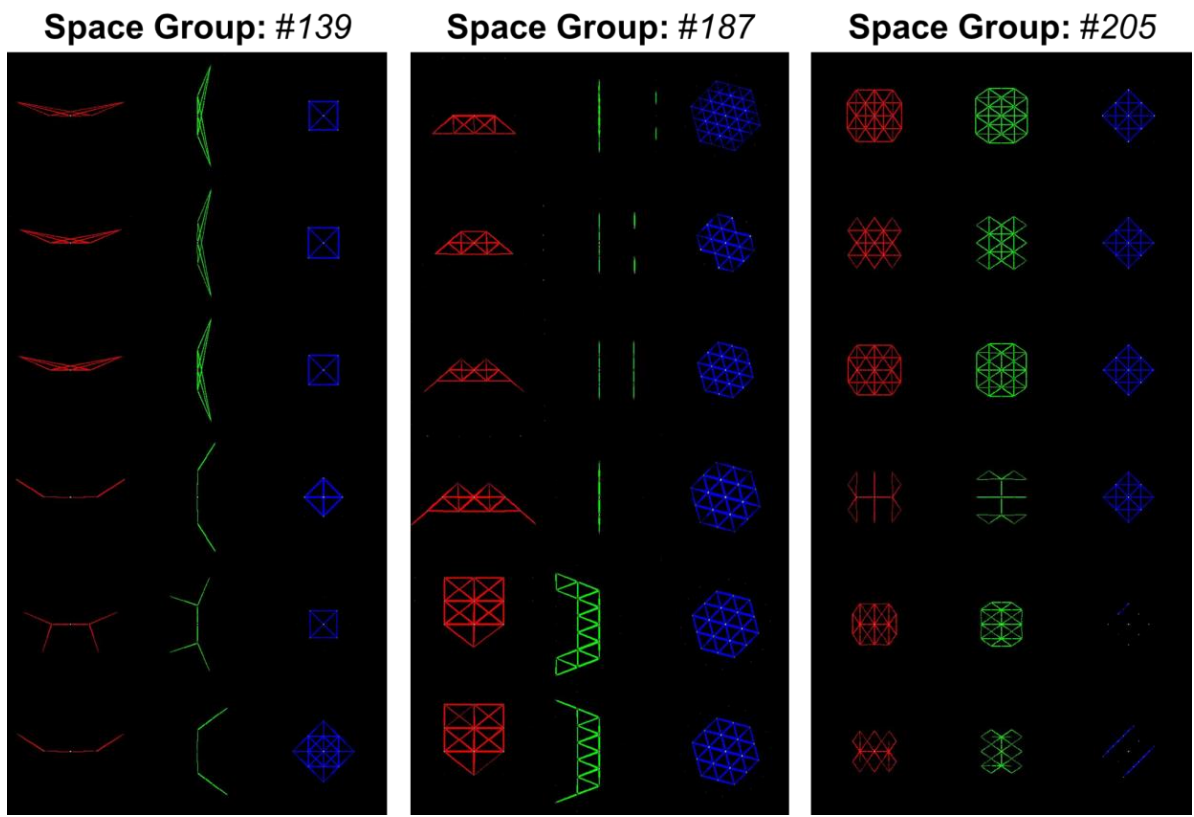

**Supplementary Figure 2. Several examples of shaped DP in three different space groups (#139, #187, and #205).**

**Supplementary Table 1. The layers' information in MSDN.**

| Network Layers | Configurations |
|---|---|
| $S_R, S_G, S_B$ | 224×224×3 |
| $conv_R^{(1)}, conv_G^{(2)}, conv_B^{(3)}$ | $3 \times 3\ conv; 2 \times 2\ avgpool$ |
| $DB_{R1}, DB_{G1}, DB_{B1}$ | $\begin{bmatrix} 1 \times 1\ conv \\ 3 \times 3\ conv \end{bmatrix} \times 3$ |
| $transition_{R1}, transition_{G1}, transition_{B1}$ | $1 \times 1\ conv; 2 \times 2\ avgpool;$ stride 2 |
| $DB_{R2}, DB_{G2}, DB_{B2}$ | $\begin{bmatrix} 1 \times 1\ conv \\ 3 \times 3\ conv \end{bmatrix} \times 6$ |
| $transition_{R2}, transition_{G2}, transition_{B2}$ | $1 \times 1\ conv; 2 \times 2\ avgpool;$ stride 2 |
| $DB_{R3}, DB_{G3}, DB_{B3}$ | $\begin{bmatrix} 1 \times 1\ conv \\ 3 \times 3\ conv \end{bmatrix} \times 8$ |
| $transition_{R3}, transition_{G3}, transition_{B3}$ | $1 \times 1\ conv; 2 \times 2\ avgpool;$ stride 2 |
| $DB_{R4}, DB_{G4}, DB_{B4}$ | $\begin{bmatrix} 1 \times 1\ conv \\ 3 \times 3\ conv \end{bmatrix} \times 6; 2 \times 2\ avgpool$ |
| $F_R, F_G, F_B$ | 1×1×960 |
| $\mathbf{\delta}_{SG}$ | 1×1×$C$ |

**Supplementary Table 2. Training and testing data information (Figures 3a-3e).**

| Crystallographic Structure | Space Group | Number of Materials |
|---|---|---|
| **Triclinic** | 1 | 7,328 |
| | 2 | 6,356 |
| **Monoclinic** | 4 | 1,430 |
| | 5 | 1,340 |
| | 6 | 1,148 |
| | 7 | 860 |
| | 8 | 2,305 |
| | 9 | 1,120 |
| | 10 | 462 |
| | 11 | 1,753 |
| | 12 | 4,324 |
| | 13 | 612 |
| | 14 | 8,592 |
| | 15 | 4,050 |
| **Orthorhombic** | 19 | 885 |
| | 25 | 330 |
| | 29 | 302 |
| | 31 | 494 |
| | 33 | 897 |
| | 36 | 622 |
| | 38 | 1,022 |
| | 44 | 392 |
| | 47 | 365 |
| | 55 | 611 |
| | 57 | 366 |
| | 58 | 394 |
| | 59 | 456 |
| | 60 | 556 |
| | 61 | 946 |
| | 62 | 5,691 |
| | 63 | 2,463 |
| | 64 | 561 |
| | 65 | 735 |
| | 70 | 313 |
| | 71 | 2,827 |
| | 74 | 946 |
| **Tetragonal** | 82 | 349 |
| | 87 | 348 |
| | 88 | 326 |
| | 107 | 317 |
| | 115 | 328 |
| | 122 | 406 |
| | 123 | 2,510 |
| | 127 | 602 |
| | 129 | 1,026 |
| | 136 | 468 |
| | 139 | 3,367 |

| Crystallographic Structure | Space Group | Number of Materials |
|---|---|---|
|  | 140 | 635 |
|  | 141 | 719 |
| **Trigonal** | 146 | 826 |
|  | 148 | 1,160 |
|  | 156 | 1,381 |
|  | 160 | 1,034 |
|  | 164 | 1,270 |
|  | 166 | 2,637 |
| **Hexagonal** | 173 | 431 |
|  | 176 | 518 |
|  | 186 | 900 |
|  | 187 | 1,094 |
|  | 189 | 997 |
|  | 191 | 803 |
|  | 193 | 583 |
|  | 194 | 3,423 |
| **Cubic** | 198 | 441 |
|  | 205 | 319 |
|  | 216 | 1,644 |
|  | 221 | 3,267 |
|  | 223 | 295 |
|  | 225 | 8,367 |
|  | 227 | 1,407 |
|  | 229 | 299 |
|  | 230 | 307 |

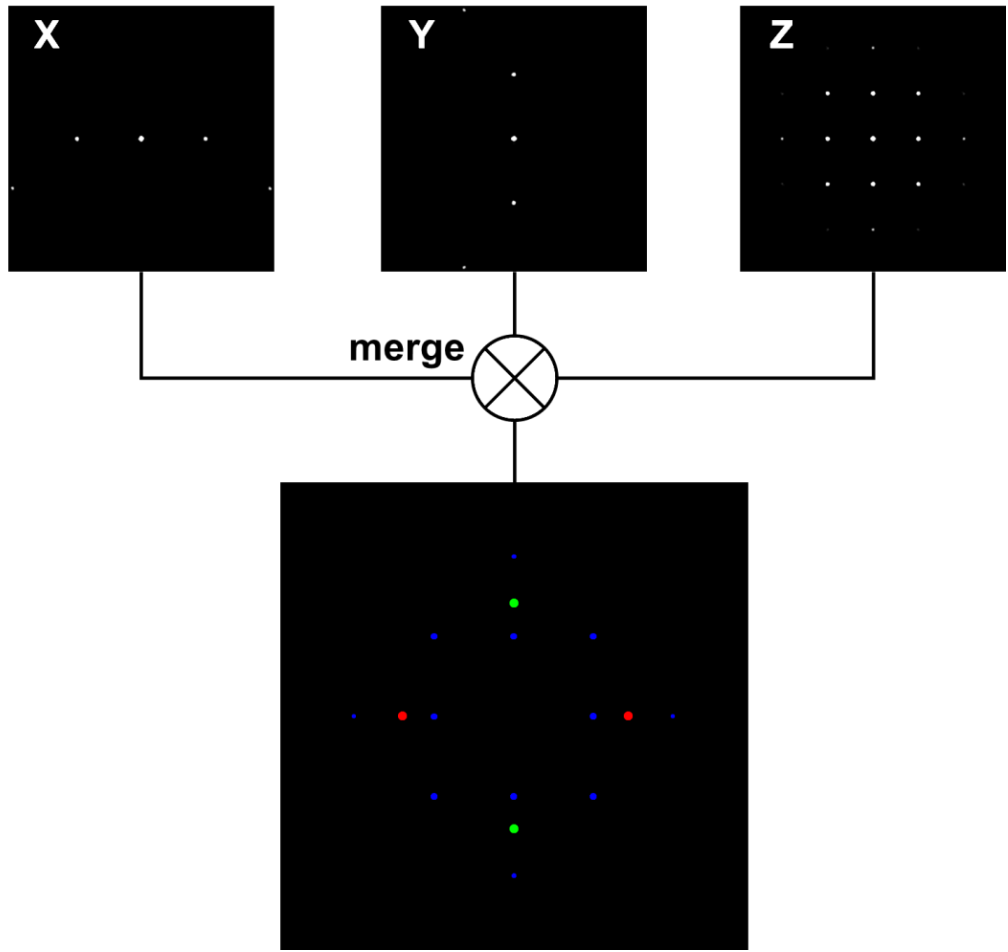

**Supplementary Figure 3. Generation process of Spot DP descriptor.** Spot DP is the superimposed (merged) version of raw DPs from RGB color channels.

**Supplementary Table 3. Test data information (Figure 3f).**

| Crystallographic Structure | Space Group | Number of Materials |
|---|---|---|
| **Monoclinic** | 3 | 198 |
| **Orthorhombic** | 69 | 146 |
| **Tetragonal** | 76 | 44 |
| | 86 | 98 |
| | 95 | 60 |
| | 102 | 46 |
| | 109 | 91 |
| | 111 | 38 |
| | 113 | 139 |
| | 128 | 142 |
| | 130 | 121 |
| | 131 | 75 |
| | 137 | 133 |
| | 142 | 190 |
| **Trigonal** | 145 | 59 |
| | 150 | 146 |
| | 162 | 155 |
| | 163 | 137 |
| | 165 | 134 |
| **Hexagonal** | 182 | 69 |
| | 185 | 152 |
| | 188 | 44 |
| **Cubic** | 197 | 45 |
| | 200 | 86 |
| | 203 | 68 |
| | 212 | 83 |
| | 214 | 52 |
| | 215 | 150 |
| | 218 | 95 |
| | 226 | 56 |

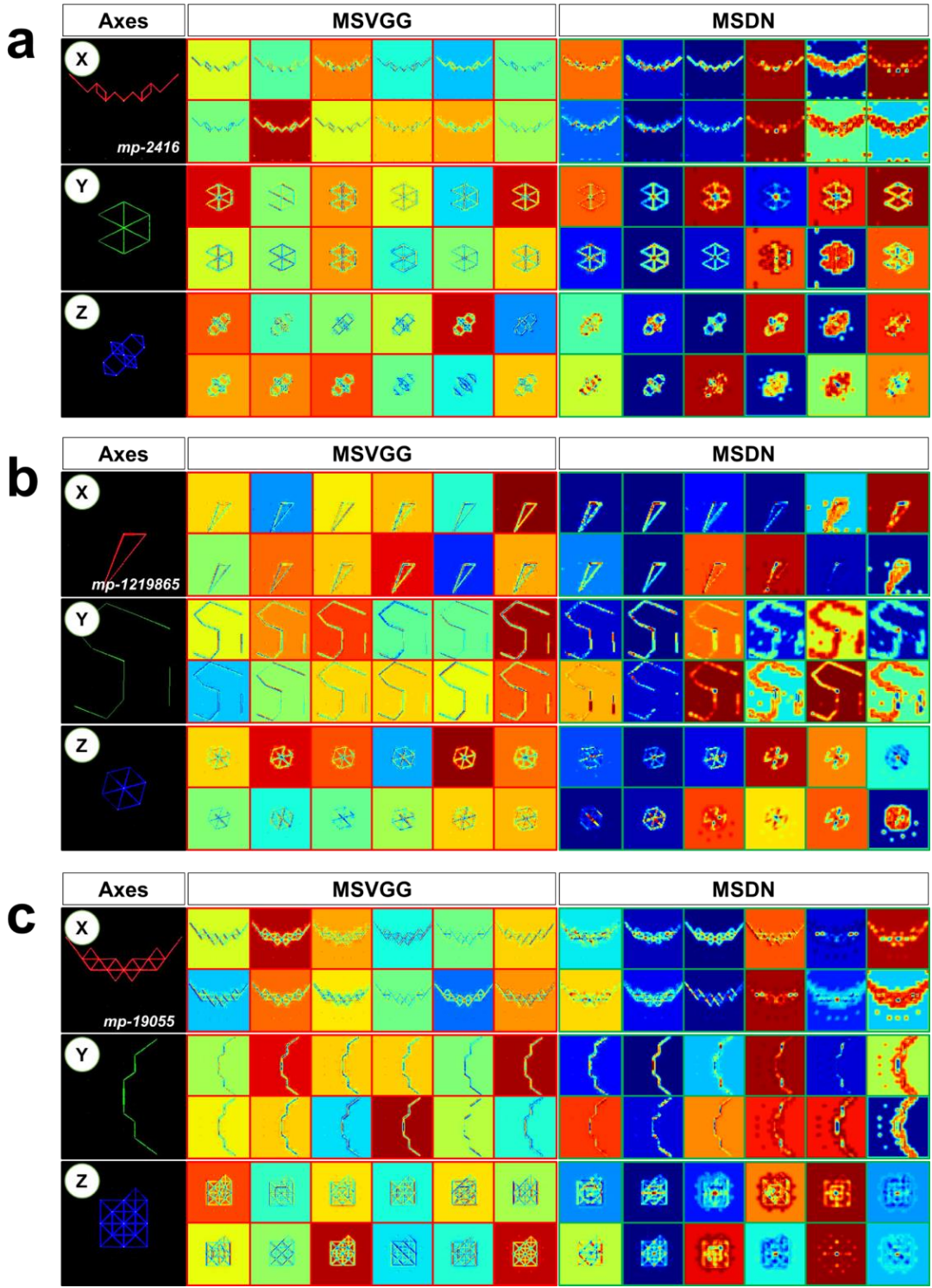

**Supplementary Figure 4. Visualization of second *conv* block in MSVGG and *DB₁* in MSDN for three selected samples (*mp-2416*, *mp-1219865*, and *mp-19055*).**

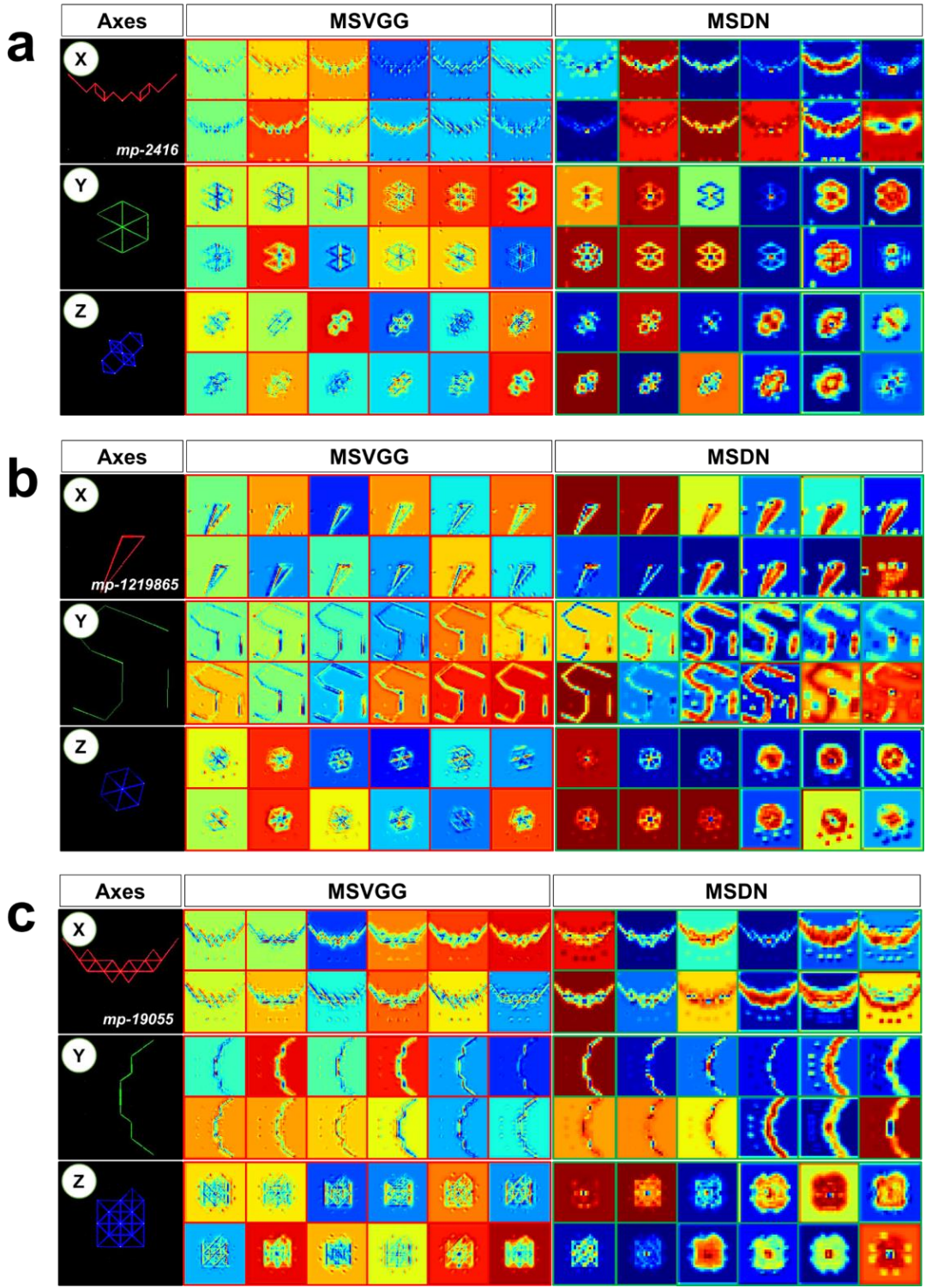

**Supplementary Figure 5.** Visualization of third *conv* block in MSVGG and *DB₂* in MSDN for three selected samples (*mp-2416*, *mp-1219865*, and *mp-19055*).

**Supplementary Table 4. The layers' information in MSVGG.**

| Network Layers | Configurations of Each Layer |
| --- | --- |
| $S_R, S_G, S_B$ | 224×224×3 |
| $conv_R^{(1)}, conv_G^{(2)}, conv_B^{(3)}$ | $f$.: 64@224×224; $k$.: 3×3; *maxpool*: 2×2 |
| $conv_R^{(4)}, conv_G^{(5)}, conv_B^{(6)}$ | $f$.: 128@128×128; $k$.: 3×3; *maxpool*: 2×2 |
| $conv_R^{(7)}, conv_G^{(8)}, conv_B^{(9)}$ | $f$.: 256@64×64; $k$.: 3×3 |
| $conv_R^{(10)}, conv_G^{(11)}, conv_B^{(12)}$ | $f$.: 256@64×64; $k$.: 3×3; *maxpool*: 2×2 |
| $conv_R^{(13)}, conv_G^{(14)}, conv_B^{(15)}$ | $f$.: 512@32×32; $k$.: 3×3 |
| $conv_R^{(16)}, conv_G^{(17)}, conv_B^{(18)}$ | $f$.: 512@32×1632 k.: 3×3; *maxpool*: 2×2 |
| $conv_R^{(19)}, conv_G^{(20)}, conv_B^{(21)}$ | $f$.: 512@16×16; $k$.: 3×3 |
| $conv_R^{(22)}, conv_G^{(23)}, conv_B^{(24)}$ | $f$.: 512@16×16; $k$.: 3×3; *maxpool*: 2×2 |
| $F_R, F_G, F_B$ | 1×1×131,072 |
| $fc_R, fc_G, fc_B$ | 1×1×4,096 |
| $\boldsymbol{\delta}_{SG}$ | 1×1×$C$ |

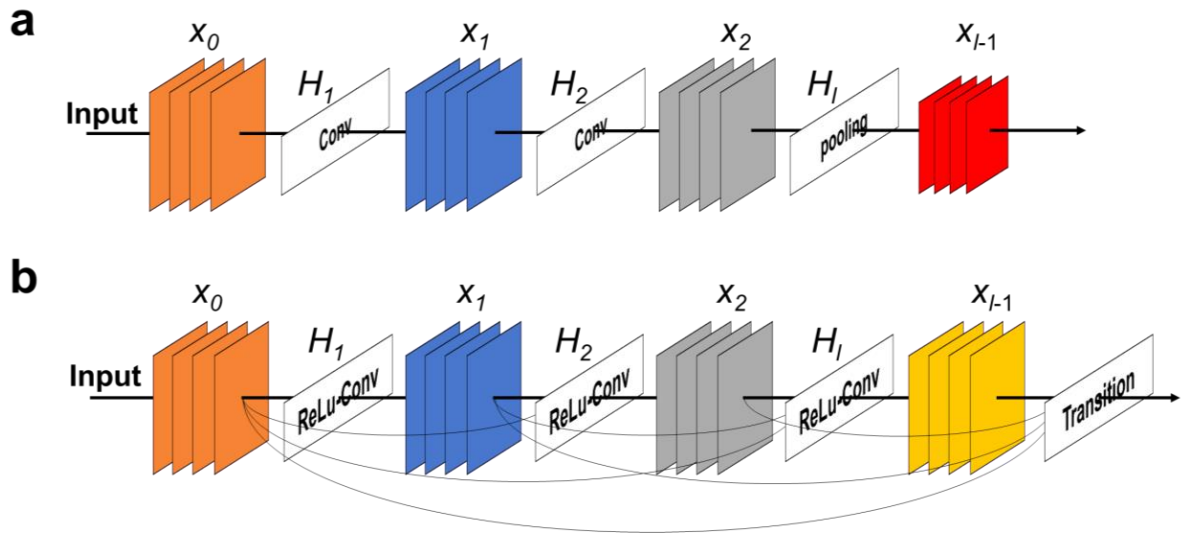

**Supplementary Figure 6. Internal structure design of MSVGG and MSDN. a-b.**, The structure of the *conv* block in MSVGG (**a**) and *DB* in MSDN (**b**).